\documentclass[12pt,letterpaper]{article}
\usepackage[margin=0.8in]{geometry}
\usepackage{amsmath}
\usepackage{abstract}
\usepackage{cite}
\usepackage{ragged2e}
\usepackage[affil-it]{authblk}
\usepackage{hyperref}
\usepackage{graphicx}
\usepackage{caption}
\usepackage{subcaption}

\begin{document}

\title{Connecting low-energy CP violation, resonant leptogenesis and neutrinoless double beta decay in a radiative seesaw model}
\author{Bikash Thapa
	    \thanks{\texttt{bikash2@tezu.ernet.in}}
	    \ and Ng. K. Francis
	    \thanks{\texttt{francis@tezu.ernet.in}}}
	    
\affil{Department of Physics, Tezpur University, Tezpur - 784028}
\date{}
\maketitle

\begin{abstract}
	We present a study of resonant leptogenesis in a radiative seesaw model. We consider the case where two quasi-degenerate right-handed neutrinos realize resonant leptogenesis, and the CP violation necessary to achieve leptogenesis occurs through the CP phases present in the neutrino mixing matrix. A numerical analysis is performed by taking the best-fit values from the current global data for three neutrino mixing angles and two mass-squared differences. We have shown how the predicted value of baryon asymmetry depends on the Dirac and Majorana CP phases. With the particular choice for the mass parameters, this model prefers a normal hierarchy of neutrino masses based on the value of baryon asymmetry predicted. Using the constrained CP phases, we evaluate the effective neutrino mass, which is relevant to the neutrinoless double beta decay.
     
\end{abstract}

\section{Introduction}
\label{sec:intro}
Evidence from neutrino oscillation experiments \cite{fukuda1998evidence} suggests that the standard model (SM) is incomplete and cannot incorporate neutrino mass. An attractive possibility of physics beyond the SM to account for tiny neutrino mass is the seesaw mechanism \cite{minkowski1977mu, yanagida1980horizontal, mohapatra1986mechanism}. In addition to explaining the tiny neutrino mass, the baryon asymmetry of the Universe (BAU) can be dynamically generated within the seesaw mechanism via leptogenesis \cite{fukugita1986barygenesis}. In leptogenesis, the CP-violating and out-of-equilibrium decays of right-handed neutrinos generate lepton asymmetry, which is then partially converted into baryon asymmetry by the electroweak sphaleron processes \cite{kuzmin1985anomalous}.

In standard thermal leptogenesis with hierarchical masses of the right-handed neutrinos, the observed value of BAU can be explained if their mass scale is $\mathcal{O}$($10^9$) GeV \cite{davidson2002lower}. However, this mass scale may be lower in cases with nearly degenerate right-handed neutrinos. Such a scenario is known as Resonant leptogenesis \cite{pilaftsis2004resonant, pilaftsis2005electroweak}. In resonant leptogenesis scenario, sufficient amount of BAU may be achieved by assuming the mass degeneracy of the right-handed neutrinos to be $\leq 10^{-8}$. Such a strong degeneracy is unnatural and fine-tuned.

 Also, since resonant leptogenesis occurs at lower temperatures, the CP phases present in the neutrino mixing matrix can act as a source of CP violation required to explain the observed BAU via leptogenesis successfully. The value of CP phases in the neutrino mass matrix is not as established as the mixing angles by various neutrino oscillation experiments. Measurements from long-baseline experiments such as T2K \cite{abe2011t2k} and NO$\nu$A \cite{ayres2005proposal} along with reactor experiments: Daya-Bay \cite{an2012observation}, RENO \cite{ahn2012observation} and Double-Chooz \cite{abe2014improved} suggests a preference for Dirac phase, $\delta \sim 1.5 \pi$. Future experiments like DUNE \cite{dune2016long} and T2HK \cite{hyper2015physics} may give a much more precise determination of the Dirac CP-violating phase. On the other hand, if neutrinos are of Majorana nature, the standard parameterization of the Pontecorvo-Maki-Naka-Sakata (PMNS) \cite{pontecorvo1958inverse, maki1962remarks, Workman:2022ynf} mixing matrix will have two additional Majorana phases. The Majorana nature of massive neutrinos may be probed in neutrinoless double beta decay experiments, and these experiments can also provide us with knowledge about the Majorana phases. These motivates us to consider the investigation of baryogenesis and explore its dependence on CP phases present in the PMNS matrix.
 
Even though the Yukawa coupling matrix, crucial to the generation of CP asymmetry through the decays of right-handed neutrinos, cannot be exclusively reconstructed by low-energy neutrino observables. However, bridging the connection between low-energy CP phases and leptogenesis is interesting. Numerous studies explore this connection in different frameworks \cite{asaka2019resonant, zhang2022leptogenesis, xing2020bridging, moffat2019leptogenesis, pascoli2007connecting, kang2021low, branco2001bridge, branco2003minimal}. The authors of reference \cite{asaka2019resonant} show the connection between low-energy CP violation and leptogenesis in the minimal seesaw model with two right-handed neutrinos. They explored the scenario of baryogenesis through resonant leptogenesis, considering all the CP violation requirement for successful leptogenesis is fulfilled by the CP phases present in the neutrino mixing matrix. Our approach is to explore the parameter space for the low-energy CP phases such that a successful explanation of the observed BAU may be given within the framework of the radiative seesaw model with an inert Higgs doublet.   
 
In this work, we consider a model of radiative neutrino masses having two right-handed neutrinos with TeV-scale masses. First, we have analyzed the scenario of resonant leptogenesis by considering a tiny splitting between the two right-handed neutrinos. An interesting feature of this model is that the observed BAU can be achieved even if the degeneracy is relaxed \cite{kashiwase2012baryon}, making the model much more natural. Then, we carry out the study of resonant leptogenesis, assuming that the CP violation required comes exclusively from the neutrino mixing matrix. In other words, we will show the parameter space for the Dirac and Majorana phases such that the observed value of BAU constrains them. 

This paper is organized as follows: In section \ref{sec:2}, we have presented a brief description of the framework of the scotogenic model with two right-handed neutrinos. We have written the Yukawa coupling matrix in Casas-Ibarra type parameterization and established assumptions that the CP violation comes exclusively from the neutrino mixing matrix. In section \ref{sec:3}, we have introduced the terms relevant for evaluating baryon asymmetry. We explore the parameter space of our model using the constraints from the observed BAU in this section. We further show the results of our numerical evaluation of the coupled Boltzmann equation, which govern the evolution of the number density of various particles involved. Section \ref{sec:4} includes the results on neutrinoless double beta decay obtained using the constrained parameter space of section \ref{sec:3}. We finally conclude our work in section \ref{sec:5}.

\section{Radiative neutrino mass model with inert an Higgs doublet}
\label{sec:2}
In this work, we have considered the study of a radiative seesaw model, first proposed by E. Ma \cite{ma2006verifiable}. The model shows an extension of the fermion sector with two right-handed neutrinos, $N_i$, and the scalar sector includes an additional inert Higgs doublet $\eta$. The particle content of the model under $SU(2)_L \times U(1)_Y \times Z_2$ may be summarised as 

\begin{align}
\label{eq:1}
	(\nu^\alpha_L, \alpha_L)  \sim  (2&, -\frac{1}{2}, +), ~~ \alpha^\alpha_R \sim (1, 1, +), ~~ (\Phi^+, \Phi^0)\sim (2, \frac{1}{2}, +)\nonumber \\ 
	& N_i \sim (1, 0, -), ~~ (\eta^+, \eta^0) \sim (2, \frac{1}{2}, -)
\end{align}

The SM fields are $Z_2$ even whereas the new fields are assumed to be odd under the $Z_2$ symmetry. The invariant Lagrangian of the new fields may be written as
\begin{equation}
\label{eq:2}
	\mathcal{L} \supset -h_{\alpha i} \bar{l}^\alpha_L \tilde{\eta} N_i + \frac{1}{2} M_i \bar{N}_i (N^c)_i + h.c.,
\end{equation}

\noindent where $h_{\alpha i}$ denote the Yukawa couplings, $l^\alpha_L$ (with $\alpha = e, \mu, \tau$) is the SM lepton doublets, $\tilde{\eta} = i \sigma_2 \eta^*$, with $\sigma_2$ being the 2$^{nd}$ Pauli matrices and $M_i$ is the Majorana masses of the right-handed neutrinos. It is assumed that the inert Higgs doublet has no vacuum expectation value ($vev$), and we find that no Dirac mass term is generated after electroweak symmetry breaking (EWSB). The neutrino remains massless at the tree level and can be generated at the one-loop level. The potential of the model becomes

\begin{align}
\label{eq:3}
	V = -\mu^2 H^\dagger H + m^2_\eta \eta^\dagger \eta + \frac{\lambda_1}{2} (H^\dagger H)^2 + \frac{\lambda_2}{2}(\eta^\dagger \eta)^2 + \lambda_3 (H^\dagger H)&(\eta^\dagger \eta) + \lambda_4 (H^\dagger \eta)(\eta^\dagger H) \nonumber \\ 
	&\frac{\lambda_5}{2}\left[(H^\dagger \eta)(H^\dagger \eta) + (\eta^\dagger H)(\eta^\dagger H)\right],
\end{align}

\noindent where $\lambda_i$ represents the quartic couplings and are assumed to be real without loss of generality. If we take $H = (0, (v+h)/\sqrt{2})^T$ and $\eta = (\eta^+, (\eta_R + i \eta_I)/\sqrt{2})^T$, the masses of the physical scalar states can be written as 

\begin{align}
\label{eq:4}
	m^2_h &= \lambda_1 v^2, \nonumber \\
	m^2_{\eta^{\pm}} &= m_\eta^2 + \frac{v^2}{2}\lambda_3, \nonumber \\
	m^2_{\eta_R} &= m_\eta^2 + \frac{v^2}{2}(\lambda_3 + \lambda_4 + \lambda_5), \nonumber \\
	m^2_{\eta_I} &= m_\eta^2 + \frac{v^2}{2}(\lambda_3 + \lambda_4 - \lambda_5),	
\end{align}

\noindent where $v$ is the $vev$ of the SM Higgs doublet, $m_h$ denotes the mass of the SM Higgs, $m_{\eta_R}$ and $m_{\eta_I}$ are the mass of the real scalar $\eta_R$ and the real pseudoscalar $\eta_I$, respectively. It is assumed that both $\eta_R$ and $\eta_I$ are lighter than the complex scalar $\eta^{\pm}$, whose mass is represented by $m^2_{\eta^{\pm}}$ in Eq. (\ref{eq:4}). It is clear that the mass difference between $\eta_R$ and $\eta_I$ is $m_{\eta_R} - m_{\eta_I} = v^2 \lambda_5$ and therefore, as $\lambda_5 \rightarrow 0$, $\eta_R$ and $\eta_I$ becomes degenerate. We note that in this model, $Z_2$ symmetry stabilizes the lightest $Z_2$ odd particle, and this particle can play the role of dark matter candidate\footnote{Here, we have restricted our study to baryon asymmetry of the Universe. A systematic study of the dark matter problem in such a model is carried out in \cite{kashiwase2012baryon, hugle2018low, honorez2007inert, dolle2009inert}}.

The radiative neutrino mass model gives rise to a neutrino mass matrix of the form\cite{ma2006verifiable}

\begin{equation}
\label{eq:5}
	\left(M_\nu\right)_{\alpha\beta} = \sum_i \frac{M_i h^*_{\alpha i} h^*_{\beta i}}{32 \pi^2} \left[L\left(m_{\eta_R}^2\right) - L\left(m_{\eta_I}^2\right)\right],
\end{equation}

\noindent with $L\left(m^2\right)$ defined as,

\begin{equation}
\label{eq:6}
	L\left(m^2\right) = \frac{m^2}{m^2 - M_i^2} \log\left(\frac{m^2}{M_i^2}\right).
\end{equation}

\noindent Analogous to the type-I seesaw formula we write the neutrino mass matrix of Eq. (\ref{eq:5}) as,

\begin{equation}
\label{eq:7}
	M_\nu = h^* \Lambda^{-1} h^\dagger,
\end{equation}

\noindent where $\Lambda$ is a diagonal matrix of the form \cite{hugle2018low},

\begin{equation}
\label{eq:8}
	\Lambda_i = \frac{2 \pi^2}{\lambda_5} \xi_i \frac{2 M_i}{v^2}
\end{equation}

\noindent and
\begin{equation}
\label{eq:9}
	\xi_i = \left(\frac{1}{8}\frac{M_i^2}{m_{\eta_R}^2 - m_{\eta_I}^2}  \left[L\left(m_{\eta_R}^2\right) - L\left(m_{\eta_I}^2\right)\right]\right)^{-1}.
\end{equation}

\noindent The active neutrino mass matrix given in Eq. (\ref{eq:5}) can be diagonalized as $U^\dagger M_\nu U^* = D_\nu = \textrm{diag}(m_1, m_2, m_3)$, where the $m_i$ are the masses of the light neutrino eigenstate. Here, $U$ is a unitary matrix also known as PMNS matrix and is represented as 

\begin{equation}
\label{eq:10}
	U = 
	\begin{pmatrix}
		c_{12}c_{13}                                      & s_{12}c_{13}                                      & s_{13} e^{-i \delta_{CP}}\\
		-c_{23}s_{12}-s_{23}c_{12}s_{13}e^{i \delta_{CP}} & c_{23}c_{12}-s_{23}s_{12}s_{13}e^{i \delta_{CP}}  & s_{23}c_{13}\\
		s_{23}s_{12}-c_{23}c_{12}s_{13}e^{i \delta_{CP}}  & -s_{23}c_{12}-c_{23}s_{12}s_{13}e^{i \delta_{CP}} & c_{23}c_{13}
	\end{pmatrix} \cdot 
	\textrm{diag}(1, e^{i\sigma}, e^{i\rho}),
\end{equation}
\noindent where $s_{ij} = \sin\theta_{ij}$ and $c_{ij} = \cos\theta_{ij}$, $\delta_{CP}$ is the Dirac phase and $\sigma$, $\rho$ are the Majorana phases. Since, we have extended the fermion sector with two right-handed neutrinos, only two light neutrinos achieve non-zero mass eigenvalues i.e., either $m_1 = 0$ (Normal Hierarchy) or $m_3 = 0$ (Inverted Hierarchy). This also means that one of the Majorana phases in $U$ becomes not well defined and only one combination of Majorana phases remains, namely, ($\sigma - \rho$) in NH or $\sigma$ in IH. 

In order to study the scenario of resonant leptogenesis in this model it is convenient to write the Yukawa matrix $h$ in Casas-Ibarra (CI) type parametrization \cite{casas2001oscillating}, 

\begin{equation}
\label{eq:11}
	h = U \sqrt{D_\nu} R^\dagger \sqrt{\Lambda}
\end{equation}

\noindent where $R$ is a complex orthogonal matrix satisfying $RR^T = 1$, which is of the form

\begin{equation}
\label{eq:12}
	R = 
	\begin{pmatrix}
		\cos \omega_{12}  & \sin \omega_{12} & 0 \\
		-\sin \omega_{12} & \cos \omega_{12} & 0 \\
		         0        &        0         & 1
	\end{pmatrix}	
	\begin{pmatrix}
		\cos \omega_{13}  & 0 & \sin\omega_{13} \\
		     0            & 1 &      0          \\
		-\sin\omega_{13}  & 0 & \cos\omega_{13}
	\end{pmatrix}
	\begin{pmatrix}
		1 &       0          &       0         \\
		0 & \cos\omega_{23}  & \sin\omega_{23} \\
		0 & -\sin\omega_{23} & \cos\omega_{23}      
	\end{pmatrix},
\end{equation}

\noindent where $\omega_{ij}$ are complex parameters. For two right-handed neutrino case, we have $\omega_{12} = \omega_{13} = 0$ in NH and $\omega_{13} = \omega_{23} = 0$ in IH case. Also, in this work we primarily focus on the possibility that the CP phases in the PMNS matrix of Eq. (\ref{eq:10}) acts as a source of the BAU. Under such a scenario we are considering Im$\omega_{23}$ or Im$\omega_{12}$ to be zero for the NH or IH case, respectively i.e., the CP violating parameters responsible for successful baryogenesis are $\delta_{CP}$ and $(\sigma-\rho)$ or $\sigma$.

\section{Resonant leptogenesis}
\label{sec:3}
In this section, we discuss the scenario of resonant leptogenesis within the radiative neutrino mass model with an inert Higgs doublet, popular as the scotogenic model. In thermal leptogenesis, where the mass spectrum of the right-handed neutrino is hierarchical, the lower bound on the mass of the right-handed neutrino is $\mathcal{O}$($10^9$) GeV \cite{davidson2002lower}. This high mass scale may be lowered in the case of resonant leptogenesis, where the masses of the right-handed neutrinos are nearly degenerate. In the type-I seesaw model with nearly degenerate right-handed neutrinos, the observed BAU can be explained through CP violating, out-of-equilibrium decays of right-handed neutrinos with TeV-scale masses. However, this comes at the cost of strong, fine-tuned degeneracy. Such a strong degeneracy is relaxed ($>10^{-8}$) in this model of a radiative seesaw with an inert Higgs doublet.

The interference of the tree-level decay of right-handed neutrino with the one-loop self-energy and vertex diagrams gives CP violation and hence produces non-zero lepton asymmetry. In resonant leptogenesis, the self-energy correction is resonantly enhanced, and the flavor-dependent asymmetry parameter is given by \cite{pilaftsis1997cp, anisimov2006cp}

\begin{equation}
\label{eq:13}
	\varepsilon_{\alpha i} = \sum_{j\neq i} \frac{\textrm{Im}\left[h^\dagger_{i\alpha} h_{\alpha j} \left(h^\dagger h\right)_{ij}\right] + \frac{M_i}{M_j} \textrm{Im}\left[h^\dagger_{i\alpha} h_{\alpha j} \left(h^\dagger h\right)_{ji}\right]}{\left(h^\dagger h\right)_{ii} \left(h^\dagger h\right)_{jj}} \cdot \frac{\left(M_i^2 - M_j^2\right) M_i \Gamma_j}{\left(M_i^2 - M_j^2\right)^2 + M_i^2 \Gamma_j^2},
\end{equation}  

\noindent where the decay width $\Gamma_i$ is defined as 

\begin{equation}
\label{eq:14}
	\Gamma_i = \frac{M_i}{8\pi}\left(h^\dagger h\right)_{ii} \left(1-\eta_i\right)^2, ~ \textrm{with}~\eta_i = \frac{m_\eta^2}{M_i^2}
\end{equation}

The CP asymmetry generated from the decay of nearly degenerate right-handed neutrinos generates the lepton asymmetry, whose value can be evaluated by analysis of the Boltzmann equation. Here, we solve the coupled Boltzmann equations (Eq. (\ref{eq:15})) which describes the evolution of the right-handed neutrino density, N$_{N_i}$ and lepton number density, N$_{N_{\alpha \alpha}}$ ($\alpha = e, \mu, \tau$) \cite{pilaftsis2004resonant, de2007quantum}.

\begin{align}
\label{eq:15}
	\frac{d\textrm{N}_{N_i}}{dz} & = -D_i\ \left(\textrm{N}_{N_i}-\textrm{N}_{N_i}^{\textrm{eq}}\right)\\
\label{eq:16}
	\frac{d\textrm{N}_{\alpha\alpha}}{dz} & = -\sum_{i=1}^2 \varepsilon_{i\alpha} D_i \left(\textrm{N}_{N_i}-\textrm{N}_{N_i}^{\textrm{eq}}\right)\nonumber \\  & \qquad -\frac{1}{4} \left\{\sum_{i=1}^2 (r z)^2 D_i  \mathcal{K}_2(r z) + W_{\Delta L=2}\right\} \textrm{N}_{\alpha\alpha},
\end{align}

\noindent The above Boltzmann equations are flavor-diagonal, and we have considered the decays, inverse decays of the right-handed neutrinos, and the $\Delta L = 2$ washout processes. The equilibrium number density of the right-handed neutrinos is given by,

\begin{equation}
\label{eq:17}
	\textrm{N}_{N_i}^\textrm{eq} = \frac{45}{2 \pi^2 g_*} z^2 \mathcal{K}_2(z),
\end{equation}
\noindent where $\mathcal{K}_2(z)$ and $g_*$ are the modified Bessel function of the second kind and the number of relativistic degrees of freedom, respectively. The decay parameter, $D_i$ is defined as

\begin{equation}
\label{eq:18}
	D_i = \frac{z}{H(z=1)}\frac{\Gamma_{N_i}}{\textrm{N}_{N_i}^\textrm{eq}},
\end{equation}

\noindent with $H$ being the Hubble parameter. The $\Delta L = 2$ washout processes consist of scattering such as $l\eta \leftrightarrow \bar{l}\eta^*$ and $ll \leftrightarrow \eta^* \eta^*$ and can be written as \cite{hugle2018low}

\begin{equation}
\label{eq:19}
	\Delta W = \frac{\Gamma_{\Delta L = 2}}{Hz} = \frac{36 \sqrt{5} M_{Pl}}{\pi^{\frac{1}{2}}g_l\sqrt{g_*}v^4} \frac{1}{z^2}\frac{1}{\lambda_5^2} M \bar{m}_\xi^2
\end{equation}
\noindent where the $\bar{m}_\xi$ is the effective mass parameter and is defined as 

\begin{align}
\label{eq:20}
	\bar{m}_\xi^2 & = \sum_{i,j} \xi_i \xi_j \textrm{Re}\left[\left(R D_\nu R^\dagger\right)_{ij}^2\right] \nonumber\\
	& \approx 4 \xi_1^2 m_1^2 + \xi_2^2 m_{2}^2 + \xi_3^2 m_3^2
\end{align}

To examine the possibility of CP phases present in the PMNS matrix as the sole source of CP violation necessary to achieve successful leptogenesis, we take Re$\omega_{23}$ or Re$\omega_{12}$ to be $\frac{\pi}{4}$ for NH or IH cases, respectively. We fix, the mass of right-handed neutrinos, $M_1 = 5$ TeV, the mass splitting parameter, $\Delta M = 10^{-7}$, $m_{\eta_R} = 1.5$ TeV, and $\lambda_5 = 10^{-3}$. We observe that with these choices of parameters, in the case of IH, the value of baryon asymmetry, $\eta_B$ is quite small, $\mathcal{O}$($10^{-20}$), and hence, we will confine our study to the NH case. Since the BAU is generated at the TeV scale, we take into consideration the flavor effects of leptogenesis. This makes the CP violation in the PMNS matrix relevant to the production of the BAU. Here, the primary focus is to show how the generated baryon asymmetry depends on the CP phases $\delta_{CP}$ and $\alpha = \sigma-\rho$.

To estimate the baryon asymmetry, $\eta_B$, we numerically solve the Boltzmann equations in Eqns. (\ref{eq:15}) and (\ref{eq:16}). We scan the $\delta_{CP}-\alpha$ plane by taking the whole range of $[0, 2\pi]$ for both $\delta_{CP}$ as well as $\alpha$ and evaluate $\eta_B$ by taking the best-fit values for the three mixing angles and two mass-squared differences from the global results of latest neutrino oscillation experiments \cite{de20212020}. To compare the numerical results for $\eta_B$ with the experimental results given by Planck \cite{ade2016planck}, we make a $\chi^2$ analysis and minimize the function,

\begin{equation}
\label{eq:21}
	\chi^2 = \sum_i \frac{\left(\lambda^{model} - \lambda^{expt}\right)^2}{\Delta \lambda^2},
\end{equation}

\noindent where $\lambda^{model}$ represents the $\eta_B$ predicted by the model, $\lambda^{expt}$ is the experimentally measured value of $\eta_B$, and $\Delta \lambda$ denote the $1\sigma$ range.

\begin{figure}[!t]
	\begin{center}
		\begin{subfigure}{0.5\textwidth}
			\includegraphics[width=\textwidth]{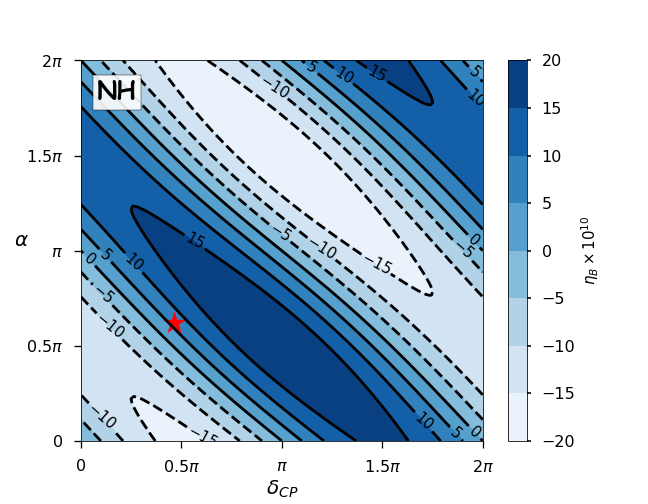}
		\end{subfigure}
		\hfill
		\begin{subfigure}{0.48\textwidth}
			\includegraphics[width=\textwidth]{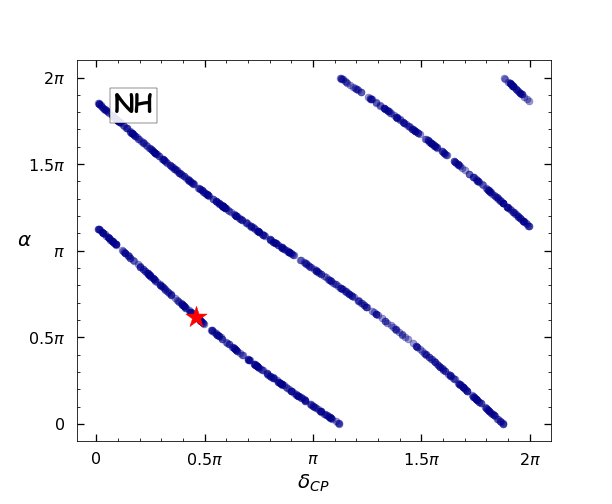}
		\end{subfigure}
		\caption{The left panel shows the contour plot of $\eta_B$ in the $\delta_{CP}-\alpha$ plane.
The red cross mark indicates the $\chi^2$-minimum value for ($\delta_{CP}, \alpha$). $\eta_B$ has negative values in regions with dotted lines. The right panel shows the allowed region of parameter space constrained by the Planck bound, $\eta_B = (6.10 \pm 0.04)\times 10^{-10}$.}
		\label{fig:1}
	\end{center}
\end{figure}

\begin{figure}[t]
	\begin{center}
		\includegraphics[width = 0.6\textwidth]{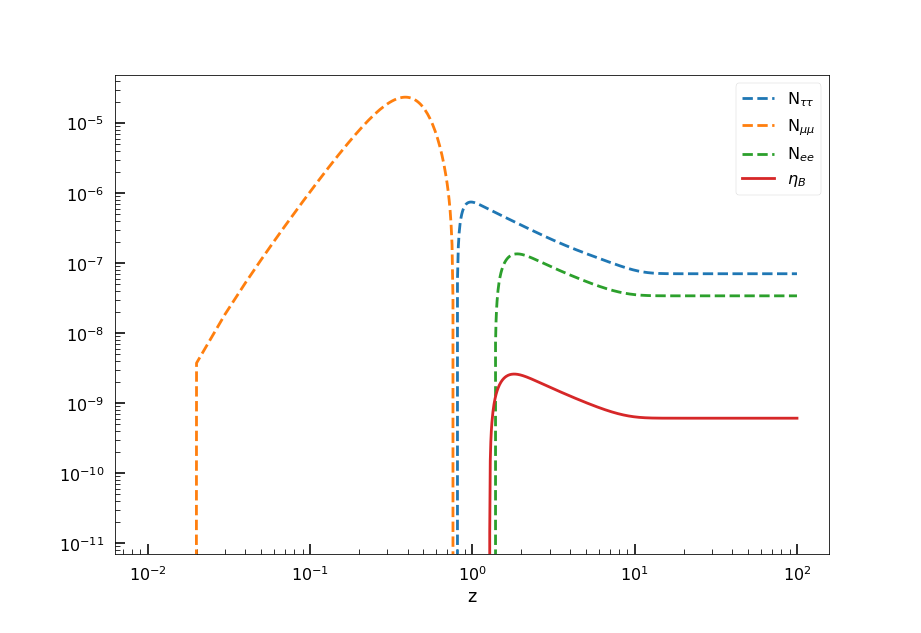}
		\caption{Variation of lepton number density N$_{\alpha\alpha}$ and $\eta_B$ as a function of $z$.}
		\label{fig:2}
	\end{center}
\end{figure}

Figure \ref{fig:1} shows the results of the parameter scan. The contour plot of $\eta_B$ in the $\delta_{CP}-\alpha$ plane is shown in the left panel. It is clear that the value of $\eta_B$ calculated in this scenario depends on both the CP phases. The right panel of figure \ref{fig:1} represents the region in $\delta_{CP}-\alpha$ space that is constrained by the observed value of BAU, $\eta_B = (6.10 \pm 0.04)\times 10^{-10}$. The best-fit value in the parameter space of our model is evaluated using Eq. (\ref{eq:21}), which is denoted by a red star mark and corresponds to $\delta_{CP} = 0.46 \pi$ and $\alpha = 0.62 \pi$. Using the best-fit point, we show in figure \ref{fig:2} the evolution of lepton number density for three flavors and baryon asymmetry as a function of $z$. The final value of baryon asymmetry is observed at a high value of $z$ and is found to be $\eta_B \approx 6.1 \times 10^{-10}$.

In this section, we demonstrate how BAU could be explained via resonant leptogenesis within the scotogenic model such that the required CP violation comes exclusively from the PMNS matrix. By confronting the predictions of our model with experimentally measured $\eta_B$, we obtained the allowed region for the CP phases. In the next section, we will study the process of neutrinoless double beta decay. We evaluate the effective light neutrino mass on which the amplitude of neutrinoless double beta decay depends. 

\section{Neutrinoless double beta decay}
\label{sec:4}
If the neutrinos are of Majorana nature it can mediate the lepton number violating neutrinoless double beta decay: $(A, Z) \rightarrow (A, Z+2) + 2e^-$ \cite{vergados2012theory, pas2015neutrinoless}. The decay rate of such a process is proportional to the effective light neutrino mass, $|\langle m_{ee}\rangle|$ and is given by \footnote{The contribution from heavy right-handed leptons is negligible, and hence we take into consideration the effect of only the light neutrinos.}

\begin{equation}
\label{eq:22}
	|\langle m_{ee}\rangle| = \lvert\sum_i m_i U_{ei}^2\rvert.
\end{equation}

\begin{figure}[!t]
	\begin{center}
		\begin{subfigure}{0.5\textwidth}
			\includegraphics[width=\textwidth]{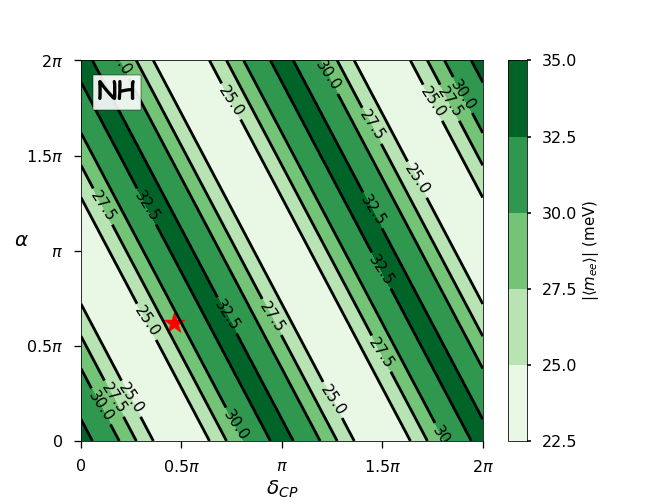}
		\end{subfigure}
		\hfill
		\begin{subfigure}{0.49\textwidth}
			\includegraphics[width=\textwidth]{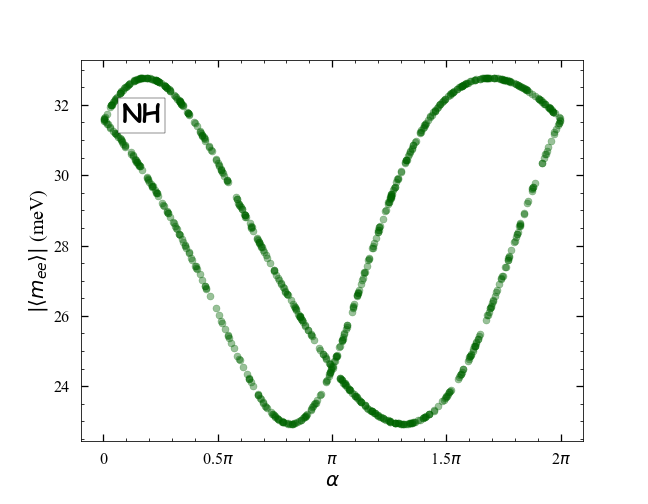}
		\end{subfigure}
		\caption{The left panel shows the contour plot of $|\langle m_{ee}\rangle|$ in $\delta_{CP}-\alpha$ plane. The red star mark indicate the best-fit point corresponding to $\chi^2$-minimum value. The right panel shows the variation of effective light neutrino mass with the Majorana phase, $\alpha$, constrained by the requirement of successful resonant leptogenesis.}
		\label{fig:3}
	\end{center}
\end{figure}

\noindent We take the allowed region of parameter space obtained through the requirement of successful resonant leptogenesis and show how the predicted $|\langle m_{ee}\rangle|$ is constrained. The left panel of figure \ref{fig:3} shows the calculated value of effective light neutrino mass, $|\langle m_{ee}\rangle|$ by scanning $\delta_{CP}$ and $\alpha$ over the range $[0, 2\pi]$. It is clear that the value of $|\langle m_{ee}\rangle|$ depends significantly on $\delta_{CP}$ as well. In the right panel of figure \ref{fig:3}, we show the predicted values of $|\langle m_{ee}\rangle|$ constrained by the successful generation of BAU via resonant leptogenesis. The predicted values range from 22.9 meV to 32.8 meV. The minimal value of $|\langle m_{ee}\rangle|$ is obtained when $\alpha = 0.8 \pi$ and the maximal value is observed at $\alpha = 1.7 \pi$.

\section{Conclusions}
\label{sec:5}
In this work, we studied a radiative seesaw model of neutrino masses with an additional inert Higgs doublet. Such a model is fascinating as it can explain the origin of neutrino masses, BAU, and Dark Matter. Here, we have considered the extension of the fermion sector by two nearly degenerate right-handed neutrinos with TeV-scale masses. Considering the quasi-degenerate nature of the right-handed neutrinos, we investigate the possibility of generating the BAU through the scenario of resonant leptogenesis. In this case leptogenesis occurs at TeV-scale temperature through the out-of-equilibrium, CP violating decay of $Z_2$ odd right-handed neutrino into $Z_2$ even SM leptons and inert Higgs doublet: $N_i\rightarrow l \eta$, $N_i\rightarrow \bar{l} \eta^*$. 

We have considered that the BAU can be generated with the CP violation, which may be measured at neutrino oscillation and neutrinoless double beta decay experiments. In other words, we have assumed that the CP violation necessary for successful leptogenesis comes from the phases present in the neutrino mixing matrix (PMNS matrix). Furthermore, we found that the model prefers NH of neutrino masses for the choices of mass parameters and the quartic coupling presented in this work. Thus, the parameters upon which the estimation of BAU depends are the Dirac phase, $\delta_{CP}$, and the Majorana phase, $\alpha = \sigma-\rho$. 

We numerically solved the coupled Boltzmann equations and found that this model can explain the observed BAU even when the source of CP violation are the CP phases, $\delta_{CP}$, and $\alpha$. Thus, we found the region of parameter space that successfully produces the observed baryon asymmetry. The best-fit point values for the parameters corresponding to $\chi^2$-minimum value are $\delta_{CP} = 0.46\pi$ and $\alpha = 0.62\pi$. Next, we discussed how the requirement of successful resonant leptogenesis constrains the effective light neutrino mass, which is relevant for neutrinoless double beta decay. Using the constrained parameter space of the model, we calculated the $\lvert\langle m_{ee}\rangle\rvert$. We found that significantly, the $\lvert\langle m_{ee}\rangle\rvert$ depends on both the CP phases. Future generation neutrinoless double beta decay experiments can probe such a model.

\section*{Acknowledgement}
BT acknowledges the Department of Science and Technology (DST), Government of India for INSPIRE Fellowship vide Grant No. DST/INSPIRE/2018/IF180588. The research of NKF is funded by DST-SERB, India under Grant No. EMR/2015/001683.
 
\bibliographystyle{unsrt}
\bibliography{references}

\end{document}